\documentclass[aps,twocolumn,prl,showpacs,superscriptaddress]{revtex4}%
\usepackage{graphicx}
\usepackage{dcolumn}
\usepackage{bm}
\usepackage{amsmath}
\usepackage{amssymb}
\usepackage{epsfig}
\usepackage{pstricks,pst-grad,pst-plot}
\usepackage{framed}
\usepackage{longtable}
\usepackage{subfigure}

\begin{document}


\title{Driving-induced crossover: from classical criticality to
  self-organized criticality}

\author{Francisco-Jos\'e P\'erez-Reche}
\affiliation{Department of Chemistry, University of
  Cambridge, Cambridge, CB2 1EW, UK}
\email{fjp23@cam.ac.uk}

\affiliation{Laboratoire de M\'ecanique des Solides, CNRS UMR-7649, Ecole
Polytechnique, Route de Saclay, 91128 Palaiseau, France}

\author{Lev Truskinovsky}
\affiliation{Laboratoire de M\'ecanique des Solides, CNRS UMR-7649, Ecole
Polytechnique, Route de Saclay, 91128 Palaiseau, France}
%
%
\author{Giovanni Zanzotto}
\affiliation{Dipartimento di Metodi e Modelli Matematici per le Scienze
Applicate,\\
Universit\`a di Padova, Via Trieste 63, 35121 Padova, Italy}

\begin{abstract}
  We propose a spin model with quenched disorder which exhibits in
  slow driving two drastically different types of critical
  nonequilibrium steady states. One of them corresponds to classical
  criticality requiring fine-tuning of the disorder. The other is a
  self-organized criticality which is insensitive to disorder. The
  crossover between the two types of criticality is determined by the
  mode of driving. As one moves from ``soft'' to ``hard'' driving the
  universality class of the critical point changes from a classical
  order-disorder to a quenched Edwards-Wilkinson universality
  class. The model is viewed as prototypical for a broad class of
  physical phenomena ranging from magnetism to earthquakes.

\end{abstract}


\maketitle

The study of criticality in externally driven inhomogeneous systems
has attracted much attention in the last two decades
\cite{Sethna}. Such systems exhibit rate independent dissipation and
are widely used to model hysteretic phenomena and intermittency
associated with magnetism, superconductivity, porous flow, fracture,
friction, plasticity and martensitic phase transitions
\cite{Kardar98}. It has been realized that in most cases scaling
emerges as an interplay between quenched disorder, extremal dynamics,
and quasi-static driving \footnote{We exclude here similar critical
  phenomena which do not require quenched disorder,
  e.g. \cite{Bak1987}.}. Within this general framework the theoretical
work has been mostly focussed on two types of models. In models of the
Random Field Ising (RFIM) type, the critical behavior requires fine
tuning of the amount of disorder $r$ and the intermittent events
(avalanches) are scale-free only at a certain $r=r_o$
\cite{Sethna}. An alternative approach links criticality to a
pinning-depinning (PD) transition where the disorder $r$ is an
irrelevant parameter \cite{Narayan1993}. It has been established that
criticality in the first class of models is classical, as in second
order phase transitions \cite{Sethna}, while in the second class it is
self tuning in the sense that infinitely slow driving brings the
system automatically on a critical manifold
\cite{Urbach1995II,Durin_review2004,Dahmen2008}.
\vskip-5pt
The two approaches are fundamentally different. The first model
describes regimes with dominating \emph{nucleation}, while the second
one deals exclusively with \emph{propagation}. It is then not
surprising that the resulting criticality is different. In the RFIM
the emerging scaling has been explained by the existence of a
classical critical point of the order-disorder (OD) type.  On the
contrary, in PD theory one encounters a range of universality classes
none of which can be formally reduced to OD. A relation between the
two types of critical phenomena has been, however,
anticipated. Previous work has shown that the presence of a nonlocal
demagnetizing field of antiferromagnetic nature (as in soft magnets)
can self tune the RFIM to display front-propagation critical exponents
\cite{Kuntz2000,Carpenter2005}. In the PD framework similar
`self-tuning', often interpreted as self organized criticality (SOC),
is obtained if the system is driven through a `weak spring' which
provides an explicit feedback mechanism \cite{Alessandro1990I}. In
this letter we use these insights to develop the first unifying model
which displays a crossover between the OD and quenched
Edwards-Wilkinson (QEW) universality classes. We show that such
crossover can be achieved experimentally by modifying the properties
of the external driving. Since the QEW model is equivalent to the Oslo
rice pile model \cite{Paczuski1996} and is therefore paradigmatic for
SOC, we are essentially dealing here with a fundamental relation
between OD and SOC.

We base our model on the observation that solids can be deformed
either by applying a force (soft device), or by controlling surface
displacements (hard device) \cite{Bertotti1998,Puglisi2000}. Both
driving mechanisms, soft and hard, can be handled simultaneously if
the system has a finite `elasticity'. To introduce this effect in a
spin setting, consider a prototypical model which we call RSSM (Random
Soft-Spin \cite{Bergman1976} or Snap-Spring \cite{Muller1977,TV04}
Model). The main difference between the RSSM and its predecessor RFIM
is the finite curvature of the energy wells and the presence of
elastic barriers. While the role of the softness of the spins is known
to be secondary in equilibrium and in the case of soft driving, where
it can be accounted for through an appropriate `dressing' of the
underlying hard-spin model \cite{Bergman1976}, the curvature of the
wells becomes crucial in the case with hard driving where it plays the
role of a regularization of the otherwise degenerate problem
\cite{Bertotti1998}.
%
%
Consider a set of $N$ bi-stable units located on the nodes $i=1,2, \dots
N$ of a cubic lattice with linear size $L=N^{1/3}$.  The state of each
snap-spring is characterized by a continuous scalar order parameter
$e_i$ measuring the local `strain'.  The energy of the system is
\begin{equation}
\label{eq:Bare_Energy}
 \phi=\frac{1}{N}\sum_{i=1}^N f_i(e_i)+\frac{1}{2N} \sum_{i,j=1}^{N} K_{ij} e_i
e_j,
\end{equation}
where $f_i(e_i)$ is a double-well potential and ${\bf K}=\{K_{ij}\}$
is the interaction kernel with sufficient rate of decay to ensure
convergence in the thermodynamic limit. In each well (one defined for
$e_i<0$ and the other for $e_i>0$) we use the approximation $f_i(e_i)=
\frac{1}{2}(e_i-s_i)^2-h'_i e_i$, where $s_i=\pm1$ is a spin variable
and $\{h'_i\}$ are random numbers representing quenched disorder.  The
system is loaded through an `elastic' device with the energy
$\phi_d=\frac{c}{2}(e-\bar{e})^2$, where $\bar{e}=(1/N)\sum_ie_i$ is
the average strain of the system of snap-springs, $c$ is the stiffness
of the loading device and $e$ is the control parameter (global
`strain'). One obtains a hard device in the limit
$c\rightarrow\infty$, and a soft device in the limit $c \rightarrow
0$, $e\rightarrow\infty$ with the stress $\sigma=ce$ fixed; the system
will be driven quasi-statically by changing the control parameter $e$,
if $c$ is finite, and $\sigma$, if $c=0$. We neglect thermal
fluctuations ($T=0$) and assume that the harmonic variables $e_i$
relax `instantaneously' and can be adiabatically eliminated. By
minimizing the total energy $\phi+\phi_d$ with respect to $e_i$ we
obtain $e_i = \tilde{e} +\sum_{j=1}^N \left(J_{ij}-k/N\right)s_j+h_i$
%
where ${\bf J}= ({\bf 1} + {\bf K})^{-1}$ is the effective interaction
between the spin variables $s_i$. We impose periodic boundary
conditions, meaning that $k_{\infty}=\sum_j J_{ij}$ does not depend on
$i$, and use the notations
$\tilde{e}=e[ck_{\infty}(ck_{\infty}+1)^{-1}]$,
$k=k_{\infty}[ck_{\infty}(1+ck_{\infty})^{-1}]$, and $h_i=\sum_jJ_{ij}
h_j'$.

Because of the absence of thermal fluctuations and to the separation
of time scales between overdamped relaxation and driving, the system
remains on a metastable branch $\{e_i(\{s_k\};e)\}$ corresponding to a
particular local minimum of the total energy until the latter ceases
to be stable \cite{TV04}. When the instability condition $s_ie_i<0$ is
reached for some $i$ the system jumps (through an avalanche) to
another locally stable branch characterized by a different spin
configuration $\{s'_i\}$. We increase $e$ by driving the system from
an initial stable configuration with $\{s_i=-1\}$ and assume that
avalanches propagate (at constant $e$) with synchronous dynamics. For
simplicity, we consider only nearest-neighbor interactions with
$J_{ii}=J_0$, $J_{ij}=J_1$, and we set $J_0=J_1=1$ so that
$k_{\infty}=7$. The renormalized disorder variables $\{h_i\}$ are
drawn from a Gaussian distribution with zero mean and standard
deviation $r$.  Under these assumptions, the RSSM becomes formally
equivalent to a nonlocal augmentation of the classical RFIM with
demagnetizing factor $k$
\cite{Sethna,Urbach1995II,Kuntz2000,Carpenter2005,Queiroz2008}.

\begin{figure}
\centering
{\includegraphics[clip=true,width=6cm]{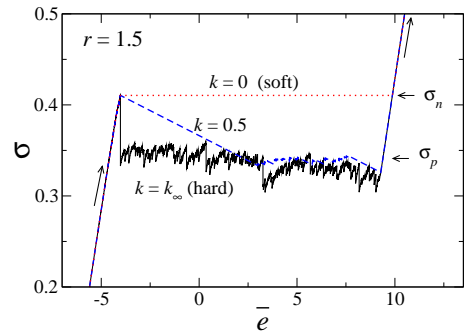}} \caption{ \label{Hysteresis}
Stress-strain  curves for
  $k=0$ (dotted line), $k=0.5$ (dashed line), and $k=k_{\infty}$
  (continuous line) for $r=1.5$ in a system with $L=64$; $\sigma_p$ and $\sigma_n$ are
  propagation and nucleation thresholds, respectively.}
\end{figure}

\begin{figure}
\centering {\includegraphics[clip=true,width=6cm]{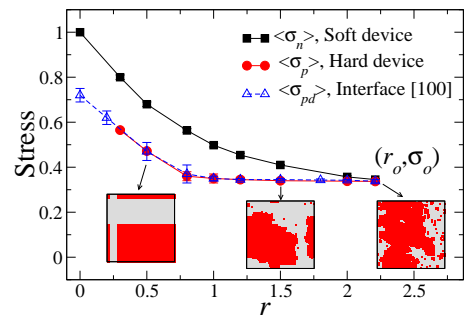}} \caption{
\label{Disorder} Dependence on disorder of the averaged nucleation and propagation
thresholds:
  $\langle \sigma_n \rangle$ (squares), $\langle \sigma_p \rangle$ at
  $k=k_{\infty}$ (circles), and $\langle \sigma_{pd} \rangle$ for an
  interface $[100]$ (triangles). Insets: cross sections of the 3D
  system showing typical transformation domains for
  $k=k_{\infty}$. Darker and lighter colors indicate transformed
  regions ($s=+1$) and untransformed regions ($s=-1$).}
\end{figure}

At $k=0$ we expectedly observe an OD transition at $r_o \simeq 2.2$
which separates a `popping' (POP) regime ($r>r_o$) where all the
avalanches are small from a `snapping' (SNAP) regime ($r<r_o$) where
an infinite avalanche sweeps most of the system
\cite{Sethna,PerezReche2003}.  The stress-strain curve $ \sigma(e) =
k_{\infty}^{-1}(\tilde{e}- kN^{-1}\sum_i s_i)$ is continuous in the
POP regime and displays a macroscopic discontinuity of the strain in
the SNAP regime [see Fig.~\ref{Hysteresis}]. The strain discontinuity
associated with the nucleation of the infinite avalanche occurs at a
`nucleation' stress $\sigma_n$ which decreases with $r$ as indicated
in Fig.~\ref{Disorder}.

For non-zero $k$, the POP regime remains essentially unaltered. In
contrast, stiffness has a non-trivial effect over the SNAP regime
observed at low disorders. In this case, a compact domain reminiscent
of the infinite avalanche (as in the $k=0$ case) nucleates at the
stress $\sigma_n$. The stress relaxes during the avalanche growth to
satisfy the global driving constraint. When $k$ is below a certain
threshold $k_p(r)$, the SNAP behavior remains qualitatively as in the
$k=0$ case because the nonlocal constraint is still soft. In contrast,
when $k>k_p(r)$, the first stress drop prevents the spanning avalanche
from growing [Fig.~\ref{Hysteresis}]. The transformation induced by
the subsequent increase of $e$ proceeds through the intermittent
growth of the previously nucleated domain with untransformed system
acting as a disordered background.
\begin{figure}
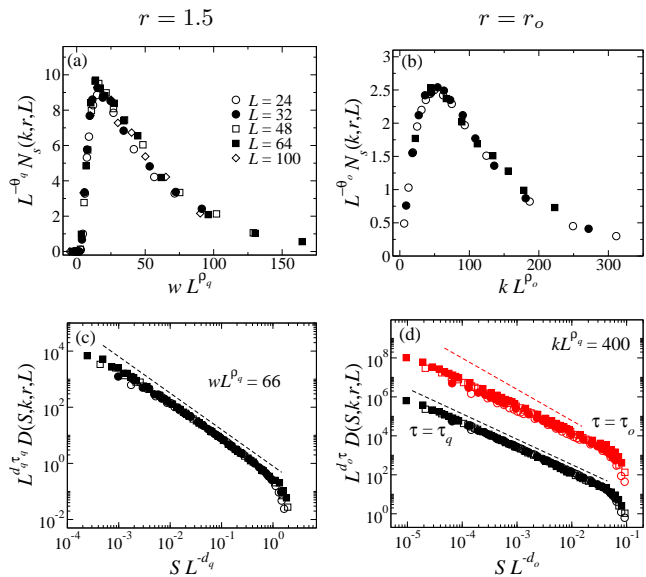

\hspace{0.3cm} $r=1.5$ \hspace{3.3cm} $r=r_o$  \\
\vspace{0.2cm} \centering
{\includegraphics[clip=true,width=4.1cm]{Ns_r1c5.eps}}
\hspace{0.03cm}
{\includegraphics[clip=true,width=4.1cm]{Ns_r2c21.eps}}\\
\vspace{0.3cm} {\includegraphics[clip=true,width=4.1cm]{Ds_r1c5.eps}}
\hspace{0.03cm}
{\includegraphics[clip=true,width=4.1cm]{Ds_r2c21.eps}}

\caption{\label{Scaling} Scaling collapse of $N_s(k,r,L)$ (a,b) and
  $D(S;k,r,L)$ (c,d) according to the hypotheses in
  Table~\ref{Table_II}. Different symbols correspond to different
  system sizes as indicated in the legend. Data in plots to the left
  (right) correspond to $r=1.5$ ($r=r_o$). Data in (c) correspond to
  $wL^{\rho_s}=66$ ; in (d) to $kL^{\rho_o}=400$. The lower scaling
  collapse in (d) is obtained with exponent $\tau_q$ while the upper
  one corresponds to $\tau_o$; dashed lines indicate the power-laws
  expected in the thermodynamic limit.}
\end{figure}
The propagation is accompanied by stress oscillations around the
propagation threshold $\sigma_p$ which remains stable during the whole
yielding process. The average of $\sigma_p$ over disorder coincides
with the depinning stress $\sigma_{pd}$ for a flat $[100]$ interface
artificially introduced and driven as in
Refs.~\cite{Roters1999,Ji1992}. This is a clear evidence that systems
with $k>k_p(r)$ reach a front-propagation regime which self-tunes
exactly around the PD point.  The surface morphology of the growing
domain is faceted at low disorders ($r \lesssim 0.8$ for systems with
$L=100$) and is self-affine at intermediate disorders.  According to
previous studies \cite{Ji1992,Bouchaud1992}, the faceted morphology is
unstable giving rise to self-affine morphology in the thermodynamic
limit for any finite $r$. Given the self-affine character of the
boundary of the nucleated domain in the case of intermediate disorders
and the short-range character of the involved interactions, one
expects to observe critical scaling of the QEW (quenched
Edwards-Wilkinson) universality class \cite{Narayan1993}.

\begin{table*}
  \caption{Scaling of the relevant quantities under a
    RG transformation with blocking parameter
    $b$ and values of the associated critical exponents (close to
    OD (subindex `\emph{o}') and to QEW (subindex `\emph{q}')).}
\begin{tabular}{lllll}
  \hline
  \hline
  &\multicolumn{2}{c}{OD} & \multicolumn{2}{c}{QEW} \\
& \multicolumn{1}{c}{RG} & \multicolumn{1}{c}{exponents} & \multicolumn{1}{c}{RG} & \multicolumn{1}{c}{exponents} \\
\hline
System size, $L$ & $L(b)=b^{-1}L$ & & $L(b)=b^{-1}L$ & \\
Stiffness, $k$ & $k(b)=b^{\rho_o} k$ & $\rho_o=1.3 \pm 0.3$ & $w(b)=b^{\rho_q} w$& $\rho_q=0.8 \pm 0.2$ \\
Disorder, $r$ & $u(b)=b^{1/\nu_o}u$& $\nu_o=1.2 \pm 0.1$ (\cite{PerezReche2003})& Irrelevant & \\
Size, $S$ & $S(b)=b^{-d_o} S$ & $d_o= 2.78 \pm 0.05$ (\cite{PerezReche2003}) & $S(b)=b^{-d_q} S$ & $d_q = 2.0 \pm 0.1$ \\
$N_s(k,r,L)$ & $N_s(b)=b^{-\theta_o} N_s$~~~ & $\theta_o=0.10 \pm 0.02$ (\cite{PerezReche2003})~~~~ & $N_s(b)=b^{-\theta_q} N_s$~~~ & $\theta_q=-0.2 \pm 0.06$\\
$D(S;k,r,L)$~~~~ & $D(b)=b^{\tau_o d_o} D$ & $\tau_o=1.6 \pm 0.06$
(\cite{Sethna}) & $D(b)=b^{\tau_q d_q} D$ & $\tau_q=1.3 \pm 0.06$ \\
\hline
$N_s(k,r,L)$ & \multicolumn{2}{c}{$L^{\theta_o} \hat{N}_s(kL^{\rho_o},ku^{-\rho_o \nu_o})$} & \multicolumn{2}{c}{$L^{\theta_q} \tilde{N}_s(wL^{\rho_q})$}\\
$D(S;k,r,L)$ & \multicolumn{2}{c}{$L^{-d_o \tau_o} \hat{D}(SL^{-d_o},
kL^{\rho_o},ku^{-\rho_o \nu_o})$} & \multicolumn{2}{c}{$L^{-d_q \tau_q}
 \tilde{D}(SL^{-d_q},wL^{\rho_q})$}\\
\hline
\hline
\end{tabular}
\label{Table_II}
\end{table*}

We check the validity of this assumption by analyzing the scaling
properties of the number of spanning avalanches $N_s(k,r,L)$ and the
distribution $D(S;k,r,L)$ of avalanche sizes $S$. Following the
arguments given in Refs.~\cite{PerezReche2003}, we consider
contributions to $N_s$ of avalanches spanning along 1 or 2 dimensions
only.  Table~\ref{Table_II} summarizes the action of the
Renormalization Group (RG) transformation on the parameters of the
system and the resulting scaling hypotheses.  The scaling variables
measuring the distance to the OD critical point are $u=r-r_o$ and
$k$. The distance to the QEW critical manifold is $w=k-k_p(r)$.
\begin{figure}

\hspace{-3.5cm}(a) \hspace{3.5cm}(b) \\


{\includegraphics[width=4.0cm,height=3.5cm,clip=true]{Phase_Diagram.eps}}
\hfill
{\includegraphics[width=4.0cm]{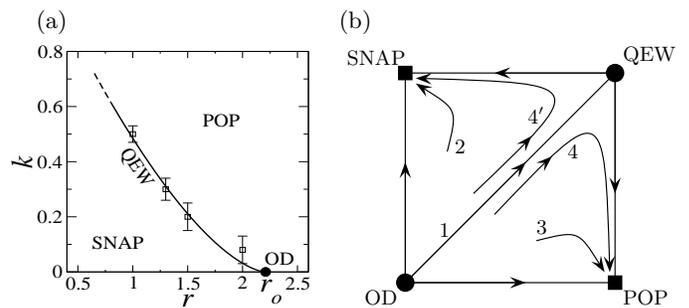}}
\caption{ \label{Phase_Diagram} (a) Schematic phase diagram for the
  RSSM (thermodynamic limit). Criticality of the OD type is associated
  with the point $r=r_o,k=0$. Criticality of the QEW type is expected
  on the line $k_p(r)$ separating SNAP and POP regimes; symbols
  display estimations of $k_p(r)$ from scaling collapses. (b)
  Schematic RG flow for the RSSM model.  Separatrix 1 going from the
  neighborhood of the OD fixed point towards the QEW fixed point
  indicates the QEW critical manifold. The RG-flow towards SNAP and
  POP regimes is indicated by arrows 2 and 3, respectively. Lines 4
  and $4'$ correspond to systems which display SOC as an intermediate
  asymptotics (QEW exponents with supercritical or subcritical cutoffs
  for $k<k_p(r)$ and $k>k_p(r)$, respectively).}
\end{figure}

Scaling collapse for $N_s(k,r,L)$ close to the line $k_p(r)$, shown in
Fig.~\ref{Scaling}(a), confirms the validity of our scaling
hypothesis. We have checked that QEW scaling persists over a finite
interval of disorders $r<r_o$. QEW scaling fails for disorders close
to $r_o$ (above $r \sim 2$ for our system sizes) due to crossover to
the OD critical regime. Fig.~\ref{Scaling}(b) displays the scaling
collapse at $u=0$ which generates the exponents $\rho_o$ and
$\theta_o$ given in Table~\ref{Table_II}.  The value of $\theta_o$
which, in contrast to $\theta_q$ is positive, agrees with previous
estimates for the RFIM \cite{PerezReche2003}.  Fig.~\ref{Scaling}(c)
presents the scaling collapse of $D(S;k,r,L)$ for $r=1.5$
corresponding to a particular section of the function
$\tilde{D}(SL^{-d_q},wL^{\rho_q})$ at constant $wL^{\rho_q}$. The
resulting exponents $d_q$ and $\tau_q$ listed in Table~\ref{Table_II}
do not depend on the selected section and are in agreement with
previous estimates from front-propagation models
\cite{Ji1992,Kuntz2000,Durin_review2004}. This confirms once again
that the propagation regime is of the QEW class; the analysis of
statistics of durations, omitted here, also supports this
interpretation. The value of $d_q$, giving the dimension of the
avalanches in the propagation regime, is consistent with $d-1$, which
confirms the self-affine morphology of the propagating domain boundary
\cite{Ji1992}. The distribution $D(S;k,r,L)$ at $r=r_0$ scales with
exponents previously reported for the OD universality class when $k$
is very small. The exponent $\tau_o$ displays a considerable crossover
to QEW when the stiffness becomes relatively large. For instance,
Fig.~\ref{Scaling}(d) shows that the scaling of $D(S;k,r,L)$ at
$kL^{\rho_o}=400$ is better with exponent $\tau_q$ than with $\tau_o$.

In Fig.~\ref{Phase_Diagram}(a) we present the phase diagram in the
$(r,k)$ plane showing stable SNAP and POP regimes separated by the QEW
line $w=0$. This line describes QEW behavior and it ends in a point
corresponding to the OD regime. The variety of observed
non-equilibrium steady states can be explained if one assumes the
existence of four fixed points for the RG flow, schematically depicted
in Fig.~\ref{Phase_Diagram}(b). The OD regime is associated with a
fully repulsive critical point which can be reached only by tuning all
four parameters: $\sigma=\sigma_c$, $r=r_o$, $k=0$, and $L^{-1}=0$. In
contrast, QEW is a saddle point with a stable manifold which governs
the large scale behavior of the systems with $r<r_o$,
$\sigma=\sigma_{pd}(r)$, $k=k_p(r)$, and $L^{-1}=0$. As we have seen
the condition $\sigma=\sigma_{pd}(r)$ is reached automatically during
the self-organized propagation regime with $k \geq k_p(r)$; the
corresponding systems lay on the critical manifold connecting OD and
QEW points. The large scale behavior for systems located away from the
critical manifold is governed either by SNAP (arrow 2) or by POP
(arrow 3) fixed points which are trivial attractors in the RG
sense. While the self-organized propagation regime is strictly
critical only for $k=k_p(r)$, our numerical simulations show that the
system exhibits truncated power law scaling with QEW exponents in a
broad range of parameters around the line $k_p(r)$ (arrows $4$ and
$4'$ in Fig.~\ref{Phase_Diagram}(b)). This observation suggests that
SOC can be viewed as an intermediate asymptotics. In general, since in
real systems stiffness is finite and disorder is generic, we
anticipate the power law structure of critical fluctuations to be in
most cases of the PD rather than of the OD type.
\vskip-2pt
We acknowledge helpful discussions with M.L. Rosinberg, S. Roux, and
G. Tarjus and partial funding from EU contract MRTN-CT-2004-505226.
\vskip-10pt
%

\begin{thebibliography}{22}
\expandafter\ifx\csname natexlab\endcsname\relax\def\natexlab#1{#1}\fi
\expandafter\ifx\csname bibnamefont\endcsname\relax
  \def\bibnamefont#1{#1}\fi
\expandafter\ifx\csname bibfnamefont\endcsname\relax
  \def\bibfnamefont#1{#1}\fi
\expandafter\ifx\csname citenamefont\endcsname\relax
  \def\citenamefont#1{#1}\fi
\expandafter\ifx\csname url\endcsname\relax
  \def\url#1{\texttt{#1}}\fi
\expandafter\ifx\csname urlprefix\endcsname\relax\def\urlprefix{URL }\fi
\providecommand{\bibinfo}[2]{#2}
\providecommand{\eprint}[2][]{\url{#2}}

\bibitem[{\citenamefont{Sethna et~al.}(2001)\citenamefont{Sethna,
      Dahmen, and Myers}}]{Sethna}
  \bibinfo{author}{\bibfnamefont{J.~P.} \bibnamefont{Sethna}},
  \bibinfo{author}{\bibfnamefont{K.~A.} \bibnamefont{Dahmen}},
  \bibnamefont{and} \bibinfo{author}{\bibfnamefont{C.~R.}
    \bibnamefont{Myers}}, \bibinfo{journal}{Nature (London)}
  \textbf{\bibinfo{volume}{410}}, \bibinfo{pages}{242}
  (\bibinfo{year}{2001}); \bibinfo{author}{\bibfnamefont{J.~P.}
    \bibnamefont{Sethna}}, \bibinfo{author}{\bibfnamefont{K.~A.}
    \bibnamefont{Dahmen}}, \bibnamefont{and}
  \bibinfo{author}{\bibfnamefont{O.}~\bibnamefont{Perkovi\'{c}}}, in
  \emph{\bibinfo{booktitle}{The science of hysteresis}} \textbf{\bibinfo{volume}{II}}, edited by
  \bibinfo{editor}{\bibfnamefont{G.}~\bibnamefont{Bertotti}}
  \bibnamefont{and} \bibinfo{editor}{\bibfnamefont{I.~D.}
    \bibnamefont{Mayergoyz}} (\bibinfo{publisher}{Academic Press},
  \bibinfo{address}{Amsterdam}, \bibinfo{year}{2006}).

\bibitem[{\citenamefont{Kardar}(1998)}]{Kardar98}
  \bibinfo{author}{\bibfnamefont{M.}~\bibnamefont{Kardar}},
  \bibinfo{journal}{Phys. Rep.} \textbf{\bibinfo{volume}{301}},
  \bibinfo{pages}{85} (\bibinfo{year}{1998});
  \bibinfo{author}{\bibfnamefont{D.~S.} \bibnamefont{Fisher}},
  \bibinfo{journal}{Phys. Rep.} \textbf{\bibinfo{volume}{301}},
  \bibinfo{pages}{113} (\bibinfo{year}{1998});
  \bibinfo{author}{\bibfnamefont{M.}~\bibnamefont{Zaiser}},
  \bibinfo{journal}{Adv.\ Phys.} \textbf{\bibinfo{volume}{55}},
  \bibinfo{pages}{185} (\bibinfo{year}{2006});
  \bibinfo{author}{\bibfnamefont{M.~J.} \bibnamefont{Alava}},
  \bibinfo{author}{\bibfnamefont{P.}~\bibnamefont{Nukala}},
  \bibnamefont{and}
  \bibinfo{author}{\bibfnamefont{S.}~\bibnamefont{Zapperi}},
  \bibinfo{journal}{Adv.\ Phys.} \textbf{\bibinfo{volume}{55}},
  \bibinfo{pages}{349} (\bibinfo{year}{2006}).

\bibitem[{\citenamefont{Narayan and Fisher}(1993)}]{Narayan1993}
  \bibinfo{author}{\bibfnamefont{O.}~\bibnamefont{Narayan}}
  \bibnamefont{and} \bibinfo{author}{\bibfnamefont{D.~S.}
    \bibnamefont{Fisher}}, \bibinfo{journal}{Phys.\ Rev.\ B}
  \textbf{\bibinfo{volume}{48}}, \bibinfo{pages}{7030 }
  (\bibinfo{year}{1993});
  \bibinfo{author}{\bibfnamefont{P.}~\bibnamefont{Chauve}},
  \bibinfo{author}{\bibfnamefont{P.}~\bibnamefont{{Le Doussal}}},
  \bibnamefont{and} \bibinfo{author}{\bibfnamefont{K.~J.}
    \bibnamefont{Wiesse}}, \bibinfo{journal}{Phys.\ Rev.\ Lett.}
  \textbf{\bibinfo{volume}{86}}, \bibinfo{pages}{1785 }
  (\bibinfo{year}{2001}).

\bibitem[{\citenamefont{Durin and Zapperi}(2006)}]{Durin_review2004}
\bibinfo{author}{\bibfnamefont{G.}~\bibnamefont{Durin}} \bibnamefont{and}
  \bibinfo{author}{\bibfnamefont{S.}~\bibnamefont{Zapperi}}, in
  \emph{\bibinfo{booktitle}{The science of hysteresis}}, \textbf{\bibinfo{volume}{II}}, edited by
  \bibinfo{editor}{\bibfnamefont{G.}~\bibnamefont{Bertotti}}
  \bibnamefont{and} \bibinfo{editor}{\bibfnamefont{I.~D.}
    \bibnamefont{Mayergoyz}} (\bibinfo{publisher}{Academic Press},
  \bibinfo{address}{Amsterdam}, \bibinfo{year}{2006}).

\bibitem[{\citenamefont{Dahmen and Ben-Zion}(2008)}]{Dahmen2008}
\bibinfo{author}{\bibfnamefont{K.~A.} \bibnamefont{Dahmen}} \bibnamefont{and}
  \bibinfo{author}{\bibfnamefont{Y.}~\bibnamefont{Ben-Zion}}, in
  \emph{\bibinfo{booktitle}{Encyclopedia of Complexity and Systems Science}}
  (\bibinfo{publisher}{Springer}, \bibinfo{year}{2008}).

\bibitem[{\citenamefont{Urbach et~al.}(1995)\citenamefont{Urbach, Madison, and
  Market}}]{Urbach1995II}
\bibinfo{author}{\bibfnamefont{J.~S.} \bibnamefont{Urbach}},
  \bibinfo{author}{\bibfnamefont{R.~C.} \bibnamefont{Madison}},
  \bibnamefont{and} \bibinfo{author}{\bibfnamefont{J.~T.}
  \bibnamefont{Market}}, \bibinfo{journal}{Phys.\ Rev.\ Lett.}
  \textbf{\bibinfo{volume}{75}}, \bibinfo{pages}{276 } (\bibinfo{year}{1995}).

\bibitem[{\citenamefont{Kuntz and Sethna}(2000)}]{Kuntz2000}
\bibinfo{author}{\bibfnamefont{M.~C.} \bibnamefont{Kuntz}} \bibnamefont{and}
  \bibinfo{author}{\bibfnamefont{J.~P.} \bibnamefont{Sethna}},
  \bibinfo{journal}{Phys.\ Rev.\ B} \textbf{\bibinfo{volume}{62}},
  \bibinfo{pages}{11699 } (\bibinfo{year}{2000}).

\bibitem[{\citenamefont{Carpenter et~al.}(2005)\citenamefont{Carpenter, Dahmen,
  Mills, and Weissman}}]{Carpenter2005}
\bibinfo{author}{\bibfnamefont{J.~H.} \bibnamefont{Carpenter}} \emph{et al.},
  \bibinfo{journal}{Phys.\ Rev.\ B} \textbf{\bibinfo{volume}{72}},
  \bibinfo{pages}{052410} (\bibinfo{year}{2005}).


\bibitem[{\citenamefont{Alessandro et~al.}(1990)\citenamefont{Alessandro,
  Beatrice, Bertotti, and Montorsi}}]{Alessandro1990I}
\bibinfo{author}{\bibfnamefont{B.}~\bibnamefont{Alessandro}} \emph{et al.},
  \bibinfo{journal}{J. Appl. Phys.} \textbf{\bibinfo{volume}{68}},
  \bibinfo{pages}{2901 } (\bibinfo{year}{1990}).


\bibitem[{\citenamefont{Cizeau et~al.}(1997)\citenamefont{Cizeau, Zapperi,
  Durin, and Stanley}}]{Cizeau1997}
\bibinfo{author}{\bibfnamefont{P.}~\bibnamefont{Cizeau}} \emph{et al.},
  \bibinfo{journal}{Phys.\ Rev.\ Lett.} \textbf{\bibinfo{volume}{79}},
  \bibinfo{pages}{4669 } (\bibinfo{year}{1997}).


\bibitem[{\citenamefont{Dickman et~al.}(2000)\citenamefont{Dickman, Mu{\~n}oz,
  Vespignani, and Zapperi}}]{Dickman2000}
\bibinfo{author}{\bibfnamefont{R.}~\bibnamefont{Dickman}} \emph{et al.},
  \bibinfo{journal}{Braz. J. Phys.} \textbf{\bibinfo{volume}{30}},
  \bibinfo{pages}{27} (\bibinfo{year}{2000}).

\bibitem[{\citenamefont{Paczuski et~al.}(1996)}]{Paczuski1996}
\bibinfo{author}{\bibfnamefont{M.}~\bibnamefont{Paczuski}}
\bibnamefont{and}
  \bibinfo{author}{\bibfnamefont{S.}~\bibnamefont{Boettcher}},
  \bibinfo{journal}{Phys.\ Rev.\ Lett.} \textbf{\bibinfo{volume}{77}},
  \bibinfo{pages}{111} (\bibinfo{year}{1996});
\bibinfo{author}{\bibfnamefont{G.}~\bibnamefont{Pruessner}},
  \bibinfo{journal}{Phys.\ Rev.\ E} \textbf{\bibinfo{volume}{67}},
  \bibinfo{pages}{030301R} (\bibinfo{year}{2003}).


\bibitem[{\citenamefont{Puglisi and Truskinovsky}(2000)}]{Puglisi2000}
  \bibinfo{author}{\bibfnamefont{G.}~\bibnamefont{Puglisi}}
  \bibnamefont{and}
  \bibinfo{author}{\bibfnamefont{L.}~\bibnamefont{Truskinovsky}},
  \bibinfo{journal}{J. Mech. Phys. Solids}
  \textbf{\bibinfo{volume}{48}}, \bibinfo{pages}{1}
  (\bibinfo{year}{2000});
  \bibinfo{author}{\bibfnamefont{E.}~\bibnamefont{Bonnot}} \emph{et al.},
  \bibinfo{journal}{Phys.\ Rev.\ B} \textbf{\bibinfo{volume}{76}},
  \bibinfo{pages}{064105} (\bibinfo{year}{2007}).

\bibitem[{\citenamefont{Bertotti}(1998)}]{Bertotti1998}
  \bibinfo{author}{\bibfnamefont{G.}~\bibnamefont{Bertotti}},
  \emph{\bibinfo{title}{Hysteresis in magnetism}}
  (\bibinfo{publisher}{Academic Press}, \bibinfo{address}{San Diego},
  \bibinfo{year}{1998});
  \bibinfo{author}{\bibfnamefont{X.}~\bibnamefont{Illa}},
  \bibinfo{author}{\bibnamefont{M.{L.~Rosinberg}}}, \bibnamefont{and}
  \bibinfo{author}{\bibfnamefont{E.}~\bibnamefont{Vives}},
  \bibinfo{journal}{Phys.\ Rev.\ B} \textbf{\bibinfo{volume}{74}},
  \bibinfo{pages}{224403} (\bibinfo{year}{2006}{\natexlab{a}});
  \bibinfo{author}{\bibfnamefont{X.}~\bibnamefont{Illa}} \emph{et al.},
  \bibinfo{journal}{Phys.\ Rev.\ B} \textbf{\bibinfo{volume}{74}},
  \bibinfo{pages}{224404} (\bibinfo{year}{2006}{\natexlab{b}}).


\bibitem[{\citenamefont{Bergman and Halperin}(1976)}]{Bergman1976}
  \bibinfo{author}{\bibfnamefont{D.~J.} \bibnamefont{Bergman}}
  \bibnamefont{and} \bibinfo{author}{\bibfnamefont{B.~I.}
    \bibnamefont{Halperin}}, \bibinfo{journal}{Phys.\ Rev.\ B}
  \textbf{\bibinfo{volume}{13}}, \bibinfo{pages}{2145 }
  (\bibinfo{year}{1976}); \bibinfo{author}{\bibfnamefont{K.~A.}
    \bibnamefont{Dahmen}} \bibnamefont{and}
  \bibinfo{author}{\bibfnamefont{J.~P.} \bibnamefont{Sethna}},
  \bibinfo{journal}{Phys.\ Rev.\ B} \textbf{\bibinfo{volume}{53}},
  \bibinfo{pages}{14872 } (\bibinfo{year}{1996}).

\bibitem[{\citenamefont{M{\"u}ller and Villaggio}(65)}]{Muller1977}
  \bibinfo{author}{\bibfnamefont{I.}~\bibnamefont{M{\"u}ller}}
  \bibnamefont{and} \bibfnamefont{P.}~\bibinfo{author}{\bibnamefont{Villaggio}},
  \bibinfo{journal}{Arch. Rat. Mech.  Anal.}
  \textbf{\bibinfo{volume}{65}}, \bibinfo{pages}{25 }
  (\bibinfo{year}{1977});
  \bibinfo{author}{\bibfnamefont{B.}~\bibnamefont{Fedelich}}
  \bibnamefont{and}
  \bibinfo{author}{\bibfnamefont{G.}~\bibnamefont{Zanzotto}},
  \bibinfo{journal}{J. Nonlinear Sci.} \textbf{\bibinfo{volume}{2}},
  \bibinfo{pages}{319 } (\bibinfo{year}{1992}).

\bibitem[{\citenamefont{Truskinovsky and Vainchtein}(2004)}]{TV04}
\bibinfo{author}{\bibfnamefont{L.}~\bibnamefont{Truskinovsky}}
  \bibnamefont{and}
  \bibinfo{author}{\bibfnamefont{A.}~\bibnamefont{Vainchtein}},
  \bibinfo{journal}{J. Mech. Phys. Solids} \textbf{\bibinfo{volume}{52}},
  \bibinfo{pages}{1421} (\bibinfo{year}{2004}); \bibinfo{author}{\bibfnamefont{G.}~\bibnamefont{Puglisi}} \bibnamefont{and}
  \bibinfo{author}{\bibfnamefont{L.}~\bibnamefont{Truskinovsky}},
  \bibinfo{journal}{J. Mech. Phys. Solids} \textbf{\bibinfo{volume}{53}},
  \bibinfo{pages}{655} (\bibinfo{year}{2005}).

\bibitem[{\citenamefont{Queiroz}(2008)}]{Queiroz2008}
\bibinfo{author}{\bibfnamefont{S.}~\bibnamefont{Queiroz}},
  \bibinfo{journal}{Phys.\ Rev.\ E} \textbf{\bibinfo{volume}{77}},
  \bibinfo{pages}{021131} (\bibinfo{year}{2008}).

\bibitem[{\citenamefont{P\'erez-Reche and
      Vives}(2003)}]{PerezReche2003}
  \bibinfo{author}{\bibfnamefont{F.~J.} \bibnamefont{P\'erez-Reche}}
  \bibnamefont{and}
  \bibinfo{author}{\bibfnamefont{E.}~\bibnamefont{Vives}},
  \bibinfo{journal}{Phys. Rev. B} \textbf{\bibinfo{volume}{67}},
  \bibinfo{pages}{134421} (\bibinfo{year}{2003}), \emph{ibid.}
  \textbf{\bibinfo{volume}{70}}, \bibinfo{pages}{214422}
  (\bibinfo{year}{2004}).

\bibitem[{\citenamefont{Roters et~al.}(1999)\citenamefont{Roters,
      Hucht, L{\"u}beck, Nowak, and Usadel}}]{Roters1999}
  \bibinfo{author}{\bibfnamefont{L.}~\bibnamefont{Roters}} \emph{et
    al.}, \bibinfo{journal}{Phys.\ Rev.\ E}
  \textbf{\bibinfo{volume}{60}}, \bibinfo{pages}{5202 }
  (\bibinfo{year}{1999}).


\bibitem[{\citenamefont{Ji and Robbins}(1992)}]{Ji1992}
  \bibinfo{author}{\bibfnamefont{H.}~\bibnamefont{Ji}}
  \bibnamefont{and} \bibinfo{author}{\bibfnamefont{M.~O.}
    \bibnamefont{Robbins}}, \bibinfo{journal}{Phys.\ Rev.\ B}
  \textbf{\bibinfo{volume}{46}}, \bibinfo{pages}{14519 }
  (\bibinfo{year}{1992});
  \bibinfo{author}{\bibfnamefont{B.}~\bibnamefont{Koiller}}
  \bibnamefont{and} \bibinfo{author}{\bibfnamefont{M.~O.}
    \bibnamefont{Robbins}}, \bibinfo{journal}{Phys.\ Rev.\ B}
  \textbf{\bibinfo{volume}{62}}, \bibinfo{pages}{5771 }
  (\bibinfo{year}{2000}).


\bibitem[{\citenamefont{Bouchaud and Georges}(1992)}]{Bouchaud1992}
  \bibinfo{author}{\bibfnamefont{J.}~\bibnamefont{Bouchaud}}
  \bibnamefont{and}
  \bibinfo{author}{\bibfnamefont{A.}~\bibnamefont{Georges}},
  \bibinfo{journal}{Phys.\ Rev.\ Lett.} \textbf{\bibinfo{volume}{68}},
  \bibinfo{pages}{3908 } (\bibinfo{year}{1992});
  \bibinfo{author}{\bibfnamefont{T.}~\bibnamefont{Emig}}
  \bibnamefont{and}
  \bibinfo{author}{\bibfnamefont{T.}~\bibnamefont{Nattermann}},
  \bibinfo{journal}{Eur. Phys. J. B} \textbf{\bibinfo{volume}{8}},
  \bibinfo{pages}{525 } (\bibinfo{year}{1999}).



\bibitem[{\citenamefont{Bak et~al.}(1987)\citenamefont{Bak, Tang, and
  Wiesenfeld}}]{Bak1987}
\bibinfo{author}{\bibfnamefont{P.}~\bibnamefont{Bak}},
  \bibinfo{author}{\bibfnamefont{C.}~\bibnamefont{Tang}}, \bibnamefont{and}
  \bibinfo{author}{\bibfnamefont{K.}~\bibnamefont{Wiesenfeld}},
  \bibinfo{journal}{Phys.\ Rev.\ Lett.} \textbf{\bibinfo{volume}{59}},
  \bibinfo{pages}{381 } (\bibinfo{year}{1987}); \bibinfo{author}{\bibfnamefont{K.}~\bibnamefont{Chen}},
  \bibinfo{author}{\bibfnamefont{P.}~\bibnamefont{Bak}}, \bibnamefont{and}
  \bibinfo{author}{\bibfnamefont{S.}~\bibnamefont{Obukhov}},
  \bibinfo{journal}{Phys.\ Rev.\ A} \textbf{\bibinfo{volume}{43}},
  \bibinfo{pages}{625} (\bibinfo{year}{1991}); \bibinfo{author}{\bibfnamefont{F.~J.} \bibnamefont{P{\'e}rez-Reche}},
  \bibinfo{author}{\bibfnamefont{L.}~\bibnamefont{Truskinovsky}},
  \bibnamefont{and} \bibinfo{author}{\bibfnamefont{G.}~\bibnamefont{Zanzotto}},
  \bibinfo{journal}{Phys.\ Rev.\ Lett.} \textbf{\bibinfo{volume}{99}},
  \bibinfo{pages}{075501} (\bibinfo{year}{2007}).

\end{thebibliography}

\end{document}